\documentclass[aps,prl,reprint,10pt,letterpaper,superscriptaddress,floatfix,nobibnotes,notitlepage]{revtex4-1}
\usepackage{textcomp}
\usepackage{amsmath}
\usepackage{amsfonts}
\usepackage{amssymb}
\usepackage{graphicx}
\usepackage{siunitx}
\usepackage{setspace}
\setcitestyle{super}
\usepackage[colorlinks, citecolor={blue}]{hyperref}

\begin{document}
\raggedbottom

\author{Yuan Cao}
\email{caoyuan@mit.edu}
\affiliation{Department of Physics, Massachusetts Institute of Technology, Cambridge, Massachusetts 02139, USA}
\author{Valla Fatemi}
\affiliation{Department of Physics, Massachusetts Institute of Technology, Cambridge, Massachusetts 02139, USA}
\author{Shiang Fang}
\affiliation{Department of Physics, Harvard University, Cambridge, Massachusetts 02138, USA}
\author{Kenji Watanabe}
\author{Takashi Taniguchi}
\affiliation{National Institute for Materials Science, Namiki 1-1, Tsukuba, Ibaraki 305-0044, Japan}
\author{Efthimos Kaxiras}
\affiliation{Department of Physics, Harvard University, Cambridge, Massachusetts 02138, USA}
\affiliation{John A. Paulson School of Engineering and Applied Sciences, Harvard University, Cambridge, Massachusetts 02138, USA}
\author{Pablo Jarillo-Herrero}
\email{pjarillo@mit.edu}
\affiliation{Department of Physics, Massachusetts Institute of Technology, Cambridge, Massachusetts 02139, USA}

\title{Magic-angle graphene superlattices: a new platform for unconventional superconductivity}

\date{\today}

\begin{abstract}

\begin{singlespace}

The understanding of strongly-correlated materials, and in particular unconventional superconductors, has puzzled physicists for decades. Such difficulties have stimulated new research paradigms, such as ultra-cold atom lattices for simulating quantum materials. Here we report on the realization of intrinsic unconventional superconductivity in a 2D superlattice created by stacking two graphene sheets with a small twist angle. For angles near \SI{1.1}{\degree}, the first `magic' angle, twisted bilayer graphene (TBG) exhibits ultra-flat bands near charge neutrality, which lead to correlated insulating states at half-filling. Upon electrostatic doping away from these correlated insulating states, we observe tunable zero-resistance states with a critical temperature $T_c$ up to \SI{1.7}{\kelvin}. The temperature-density phase diagram shows similarities with that of the cuprates, including superconducting domes. Moreover, quantum oscillations indicate small Fermi surfaces near the correlated insulating phase, in analogy with under-doped cuprates. The relative high $T_c$, given such small Fermi surface (corresponding to a record-low 2D carrier density of $\sim$\SI[retain-unity-mantissa=false]{1e11}{\per\centi\meter\squared}), renders TBG among the strongest coupling superconductors, in a regime close to the BCS-BEC crossover. These novel results establish TBG as the first purely carbon-based 2D superconductor and as a highly tunable platform to investigate strongly-correlated phenomena, which could lead to insights into the physics of high-$T_c$ superconductors and quantum spin liquids.

\end{singlespace}

\end{abstract}

\maketitle

Strong interactions among particles lead to fascinating states of matter, including quark-gluon plasma, the various forms of nuclear matter within neutron stars, strange metals, and fractional quantum Hall states. \cite{rajagopal2000, strangemetal, stormer1999} An intriguing class of strongly-correlated materials are unconventional superconductors, which range from heavy-fermion and organic superconductors with relatively low critical temperature $T_c$, to iron pnictides and cuprates that can have $T_c$ in excess of \SI{100}{\kelvin}.\cite{pfleiderer2009,ishiguro1998,lee2006,keimer2015,stewart2011} Despite very intense experimental effort to characterize these materials, unconventional superconductors pose a formidable challenge to theory. Such difficulties have motivated the development of alternative approaches for investigating and modeling strongly correlated systems. One route is to simulate quantum materials with ultra-cold atoms trapped in optical lattices, \cite{bloch2008} though realizing $d$-wave superfluidity with ultra-cold atoms still awaits technical breakthroughs to reach lower temperatures\cite{mazurenko2017}. In this article, we report the observation of unconventional superconductivity in a completely new platform ------ a 2D superlattice made from graphene, namely `Magic' Angle Twisted Bilayer Graphene (MA-TBG). Created by the moir\'{e} pattern between two graphene sheets, the MA-TBG superlattice has a periodicity of about \SI{13}{\nano\meter}, in between that of crystalline superconductors (a few angstroms) and optical lattices (about a micrometre). One of the key advantages of this system is the \emph{in situ} electrical tunability of the charge carrier density in an ultra-flat band with a bandwidth on the order of \SI{10}{\milli\electronvolt}, allowing us to study the phase diagram of unconventional superconductivity in unprecedented resolution, without relying on multiple devices possibly hampered by different disorder realizations. The observed superconductivity shows several features similar to cuprates, including dome structures in the phase diagram and quantum oscillations that point towards small Fermi surfaces near a correlated insulator state. Furthermore, the observed superconductivity occurs for record-low carrier densities of a few \SI[retain-unity-mantissa=false]{1e11}{\per\centi\meter\squared}, orders of magnitude lower than typical 2D superconductors. The relatively high $T_c=\SI{1.7}{\kelvin}$ for such small densities puts MA-TBG among the strongest coupling superconductors, in the same league as cuprates and recently identified FeSe thin layers\cite{wang2017}. Our results also establish MA-TBG as the first purely carbon-based 2D superconductor and, more importantly, as a relatively simple and highly tunable platform that enables thorough investigation of strongly correlated physics.

\begin{figure*}
\includegraphics[width=0.7\textwidth]{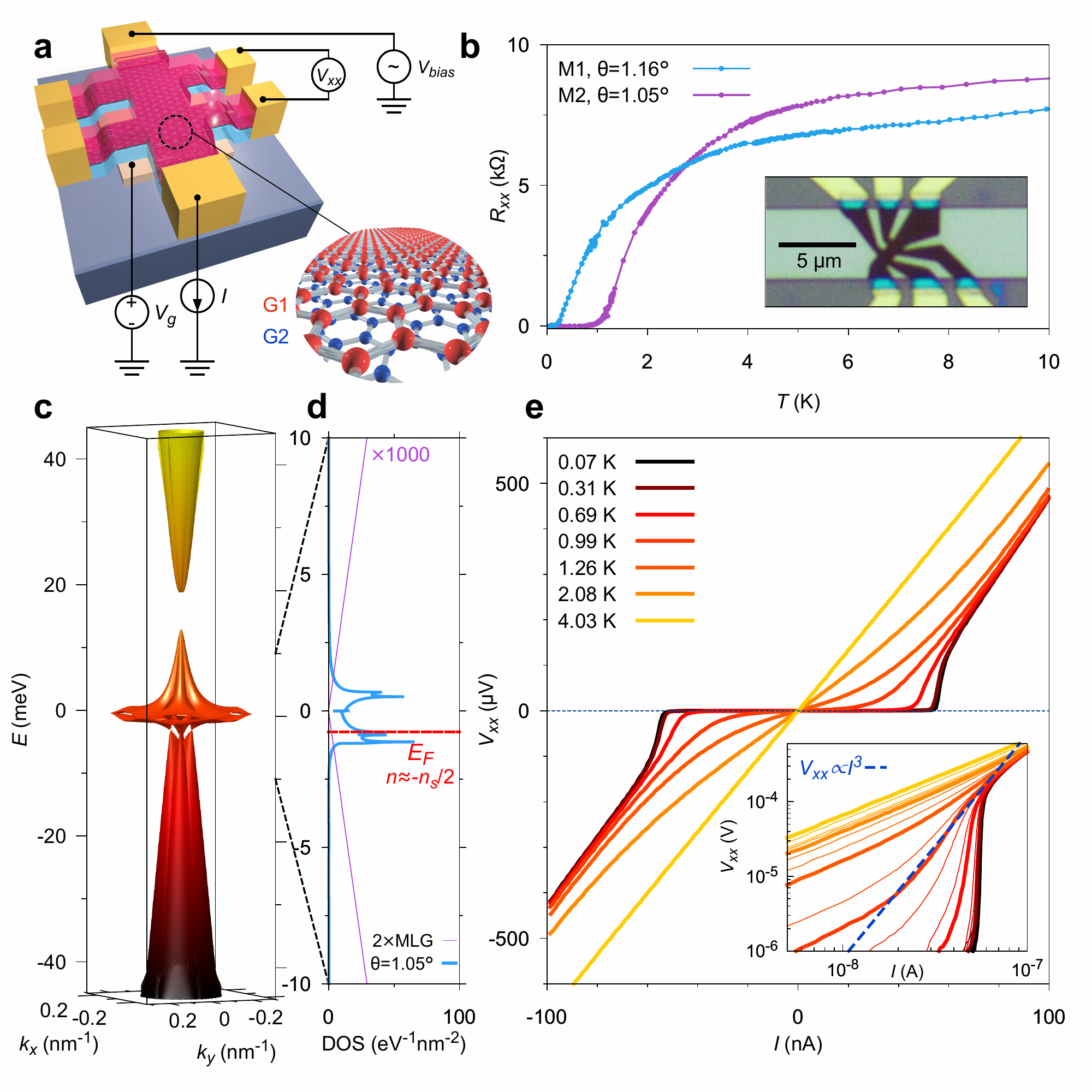}
\caption{\textbf{2D superconductivity in a graphene superlattice.} (a) Schematic of a typical twisted bilayer graphene (TBG) device and four-probe measurement scheme. The stack consists of top hexagonal boron nitride (h-BN), rotatated graphene bilayers (G1, G2) and bottom h-BN. The electron density is tuned by a metal gate beneath the bottom h-BN. (b) Measured four-probe resistance $R_{xx}=V_{xx}/I$ ($V_{xx}$ and $I$ defined in (a)) in two devices M1 and M2, with twist angles $\theta=\SI{1.16}{\degree}$ and $\theta=\SI{1.05}{\degree}$ respectively. The inset is the optical image of device M1, showing the main Hall bar (dark brown), electrical contact (gold), back gate (light green) and SiO\textsubscript{2} substrate (dark gray). (c) The band structure of TBG at $\theta=\SI{1.05}{\degree}$ in the first mini-Brillouin zone (MBZ) of the superlattice. The bands near charge neutrality ($E=0$) have an energy scale of $<\SI{15}{\milli\electronvolt}$. (d) The density of states (DOS) corresponding to the bands shown in (c), zoomed in to -10 to \SI{10}{\milli\electronvolt}. For comparison, the purple lines show the DOS of two sheets of freestanding graphene without interlayer interaction (multiplied by $1000$). The red dashed line shows the Fermi energy at half-filling of the lower branch ($E<0$) of the flat bands, which corresponds to a density $n=-n_s/2$ where $n_s$ is the superlattice density defined in the main text. The superconductivity is observed near this half-filled state. (e) $I-V$ curves for device M2 measured at $n=\SI{-1.47e12}{\per\centi\meter\squared}$ and various temperatures. At the lowest temperature of \SI{70}{\milli\kelvin}, the $I-V$ curve shows a critical current of approximately \SI{50}{\nano\ampere}. The inset shows the same data plotted in log-log scale, typically used to extract a Berezinskii-Kosterlitz-Thouless (BKT) transition temperature ($T_\mathrm{BKT}=\SI{1.0}{\kelvin}$ in this case), obtained by fitting to a $V_{xx}\propto I^3$ power law.}
\end{figure*}

Monolayer graphene has a linear energy dispersion at its charge neutrality point. When two aligned graphene sheets are stacked, the hybridization of their bands due to interlayer hopping results in fundamental modifications to the low-energy band structure depending on the stacking order (AA or AB stacking). If an additional twist angle is present, a hexagonal moir\'{e} pattern consisting of alternating AA- and AB-stacked regions emerges and acts as a superlattice modulation. \cite{morell2010, bistritzer2011, laissardiere2012, moon2012, fang2016} The superlattice potential folds the band structure into the mini Brillouin zone (MBZ). Hybridization between adjacent Dirac cones in the MBZ has an effect on the Fermi velocity at the charge neutrality point which is reduced from the typical value of \SI{1e6}{\meter\per\second}. \cite{cao2016, cao2018, morell2010, bistritzer2011, laissardiere2012, moon2012, fang2016}At low twist angles, each electronic band in the MBZ has a four-fold degeneracy of spins and valleys, the latter of which are inherited from the original graphene electronic structure.\cite{cao2016,bistritzer2011,santos2012} For convenience, we define the superlattice density $n_s=4/A$ to be the density corresponding to full-filling of each set of degenerate superlattice bands, where $A\approx\frac{\sqrt{3}a^2}{2\theta^2}$ is the moir\'{e} unit cell area, $a=\SI{0.246}{\nano\meter}$ is the lattice constant of the underlying graphene lattice, and $\theta$ is the twist angle. In the supplementary video, we present an animation of how the band structure in the MBZ of TBG evolves from $\theta=\SI{3}{\degree}$ to $\theta=\SI{0.8}{\degree}$, calculated using the continuum model for one valley.\cite{bistritzer2011}

Special angles, namely the `magic angles', exist where the Fermi velocity drops to zero\cite{bistritzer2011}, the first of which is about $\theta_\mathrm{magic}=1.1^\circ$. Near this twist angle, the energy bands near charge neutrality, which are separated from other bands by single-particle gaps, become remarkably flat. The typical energy scale for the entire bandwidth is about \SIrange{5}{10}{\milli\electronvolt} (Fig. 1c).\cite{bistritzer2011,cao2018}  Experimentally confirmed consequences of the flatness of these bands are high effective mass in the flat bands (as observed in quantum oscillations), and correlated insulating states at half-filling of these bands corresponding to $n=\pm n_s/2$, where $n=CV_g/e$ is the charge density defined by the gate voltage $V_g$ ($C$ is the gate capacitance).\cite{cao2018} These insulating states are understood as a result of the competition between Coulomb energy and quantum kinetic energy, which gives rise to a correlated insulator at half-filling and which exhibits characteristics consistent with a Mott-like insulator behavior. \cite{cao2018} The doping density required to reach the Mott-like insulating states is $n_s/2\approx 1.2\sim\SI{1.6e12}{\per\centi\meter\squared}$, depending on the exact twist angle. In this article, we show transport data which clearly demonstrate that superconductivity is achieved as one dopes slightly away from the Mott-like insulating state in MA-TBG. We have observed superconductivity across multiple devices with slightly different twist angles, with the highest critical temperature achieved of \SI{1.7}{\kelvin}.

\section*{Superconductivity in MA-TBG}

Figure 1a shows the typical device structure of fully encapsulated TBG devices. The top and bottom pieces of graphene originate from the same exfoliated flake, which allows for a relative twist angle that is precisely controlled to within \SIrange{0.1}{0.2}{\degree}.\cite{cao2016,kim2016,kim2017} The encapsulated TBG stack is etched into a Hall-bar shape and contacted from the edges.\cite{wang2013} Electrical contacts are made from non-superconducting materials (thermally evaporated Au on a Cr sticking layer), to avoid any potential proximity effect. The carrier density $n$ is tuned by applying voltage to a Pd/Au bottom gate electrode. Figure 1b shows the longitudinal resistance $R_{xx}$ as a function of temperature for two magic-angle devices, M1 and M2, with twist angles of \SI{1.16}{\degree} and \SI{1.05}{\degree} respectively. At the lowest temperature of \SI{70}{\milli\kelvin}, both devices show zero resistance, and thus a superconducting state. The critical temperature $T_c$ as calculated from the 50\% normal state resistance value is approximately \SI{1.7}{\kelvin} and \SI{0.5}{\kelvin} for the two devices we studied in detail. Figure 1c and 1d show a calculated single-particle band structure and density of states (DOS) near the charge neutrality point for $\theta=\SI{1.05}{\degree}$. The superconductivity in both devices occurs when the Fermi energy is tuned away from charge neutrality ($E_F=0$) to be near half-filling of the lower flat band ($E_F<0$, as indicated in Fig. 1d). The DOS within the energy scale of the flat bands is more than 3 orders of magnitudes higher than that of two uncoupled graphene sheets due to the reduction of Fermi velocity and increase of localization that occurs near the magic angle. We note, however, that the energy where the DOS peaks does not generally coincide with the density required to half-fill the bands. We also did not observe any appreciable superconductivity when the Fermi energy is tuned into the flat conduction bands ($E_F>0$). Figure 1e shows the $I$-$V$ curves of device M2 at different temperatures. We observe typical behavior for a 2D superconductor. The inset shows a tentative fit of the same data to a $V_{xx}\propto I^3$ power law which is predicted in a Berezinskii-Kosterlitz-Thouless (BKT) transition in 2D superconductors\cite{tinkham1996}. The use of this analysis gives a BKT transition temperature of approximately $T_\mathrm{BKT}=\SI{1.0}{\kelvin}$ at $n=\SI{-1.44e12}{\per\centi\meter\squared}$, where as before, $n$ is the charge density induced by the gate and measured from the charge neutrality point (which is different than the actual carrier density involved in transport, as we show below).

\begin{figure*}
\includegraphics[width=\textwidth]{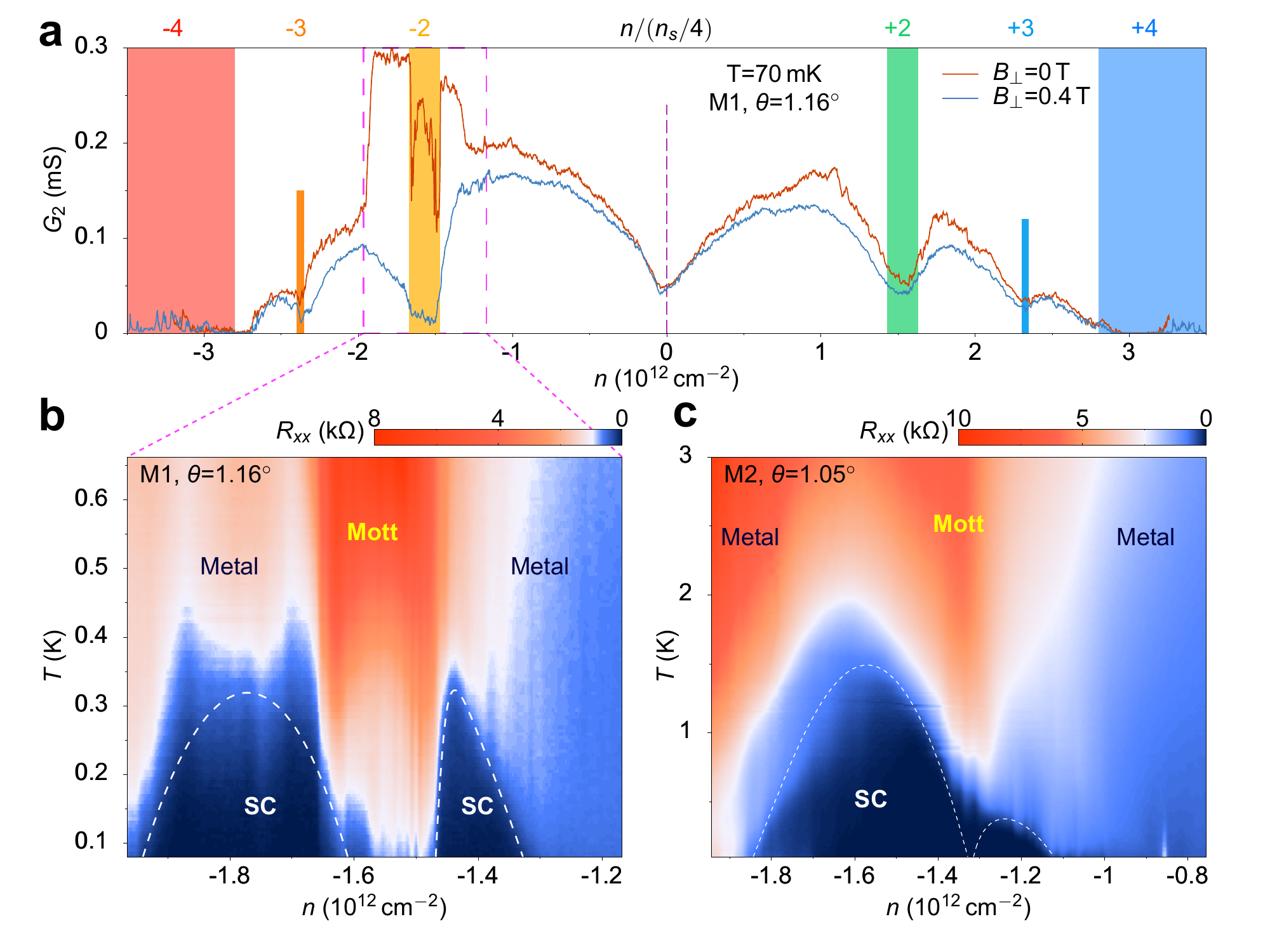}
\caption{\textbf{Gate-tunable superconductivity in MA-TBG.} (a) Two-probe conductance $G_2=I/V_\mathrm{bias}$ of device M1 measured in zero magnetic field (red trace) and at a perpendicular field of $B_\perp=\SI{0.4}{\tesla}$ (blue trace). The traces show the typical V-shaped conductance near charge neutrality $n=0$, as well as insulating states at the superlattice bandgaps, $n=\pm n_s$, corresponding to filling $\pm4$ electrons in each moir\'{e} unit cell, and the conductance reductions at intermediate integer fillings of the superlattice due to Coulomb interactions. Near $-2e^-$ per unit cell filling, there is a considerable conductance enhancement which is suppressed in $B_\perp=\SI{0.4}{\tesla}$, signaling the onset of superconductivity. Measurements are taken at $T=\SI{70}{\milli\kelvin}$. (b) Four-probe resistance $R_{xx}$ measured at densities corresponding to the region bounded by pink dashed lines in (a), versus temperature. Two superconducting (SC) domes are clearly observed next to the half-filling state (``Mott", centered around $-n_s/2=\SI{-1.58e12}{\per\centi\meter\squared}$). The remaining regions in the diagram are labeled as ``Metal" due to the metallic temperature dependence. The highest critical temperature observed in device M1 is $T_c=\SI{0.5}{\kelvin}$ (50\% normal state resistance). (c) Similar plot as in (b) but measured in device M2, showing two asymmetric and overlapping domes. The highest critical temperature in this device is $T_c=\SI{1.7}{\kelvin}$.}
\end{figure*}

In contrast to other known 2D and layered superconductors, the superconductivity in MA-TBG only requires the application of a small gate voltage, corresponding to a density of merely \SI{1.2e12}{\per\centi\meter\squared} from charge neutrality, an order of magnitude lower compared to \SI{1.5e13}{\per\centi\meter\squared} in LaAlO\textsubscript{3}/SrTiO\textsubscript{3} interfaces and \SI{7e13}{\per\centi\meter\squared} in electrochemically doped MoS\textsubscript{2}, among others.\cite{yu2016} Therefore, gate-tunable superconductivity can be realized in a high-mobility system without the need of ionic-liquid gating or chemical doping. Figure 2a shows the two-probe conductance of device M1 versus $n$ at zero magnetic field and at \SI{0.4}{\tesla} perpendicular magnetic field. Near the charge neutrality point $n=0$, a typical V-shape conductance is observed which originates from the renormalized Dirac cones of the TBG band structure. The insulating states centered at approximately $\pm\SI{3.2e12}{\per\centi\meter\squared}$ (which is $n_s$ for $\theta=\SI{1.16}{\degree}$) are due to single-particle band gaps in the band structure, which correspond to filling $\pm4$ electrons in each superlattice unit cell. In between, there are conductance minima at $\pm2$ and $\pm3$ electrons per unit cell. These are understood as many-body gaps induced by the competition between Coulomb energy and the reduced kinetic energy due to confinement of the electronic state in the superlattice near the magic angle, giving rise to insulating behavior near these integer fillings.\cite{cao2018} A possible mechanism for the gaps is thought to be similar to a Mott insulator,\cite{mott1990,imada1998} but with an extra two-fold degeneracy (for the case of $\pm2$ electrons) from the valleys in the original graphene Brillouin zone.\cite{cao2016, cao2018} In the vicinity of -2 electrons per unit cell ($n=-1.3$ to \SI{-1.9e12}{\per\centi\meter\squared}) and at a temperature of \SI{70}{\milli\kelvin}, the conductance is substantially higher at zero magnetic field than in a perpendicular magnetic field $B_{\perp}=\SI{0.4}{\tesla}$, consistent with mean-field suppression of a superconducting state by the magnetic field. Here, the maximum conductance is only limited by the contact resistance, which is absent in the four-probe measurements shown in the subsequent figures.

Figure 2b and 2c show the four-probe resistance of device M1 and M2, respectively, as a function of both density $n$ and temperature $T$. Both devices show two pronounced superconducting `domes' on each side of the half-filling correlated insulating state. These features share similarities with the phenomenology observed in high-temperature superconductivity in cuprate materials. At base temperature, the resistance inside the domes is lower than our measurement noise floor, which is more than 2 and 3 orders of magnitude lower than the normal state resistance for device M1 and M2, respectively. The $I$-$V$ curves inside the domes show critical current behavior as exemplified in Fig. 1e, while being ohmic in the metallic phases outside the domes.  When cooling down right through the middle of the half-filling state, the correlated insulating phase is exhibited at intermediate temperatures (from \SI{1}{\kelvin} to \SI{4}{\kelvin}), but at lower temperatures both devices exhibit signs of superconductivity at the lowest temperatures. Device M1 becomes weakly superconducting, while device M2 becomes fully superconducting. This may be explained by a coexistence of superconducting and insulating phases due to sample inhomogeneity. 

\section*{Magnetic field response}

\begin{figure*}
\includegraphics[width=\textwidth]{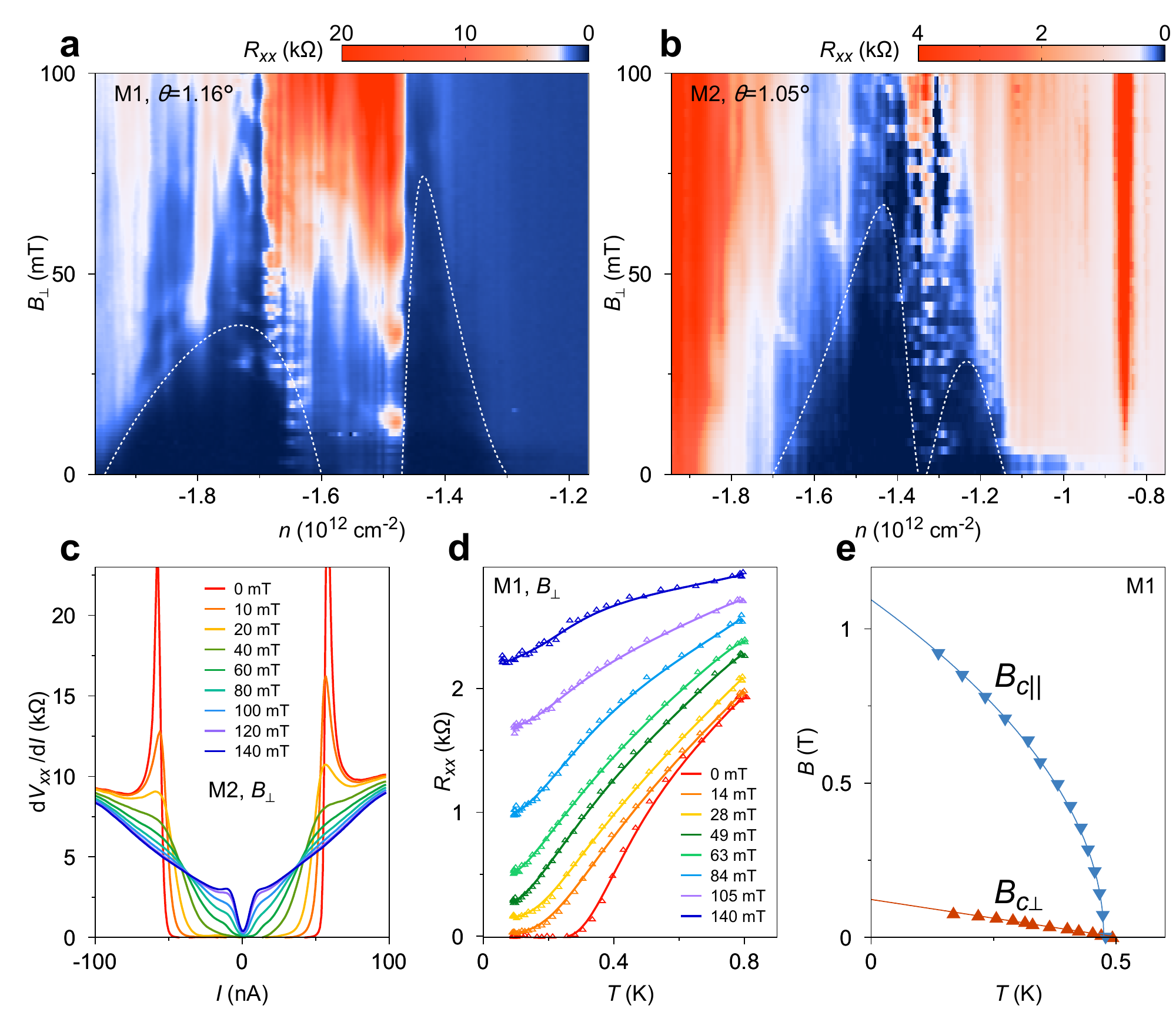}
\caption{\textbf{Magnetic field response of the superconducting states in MA-TBG.} (a-b) Four-probe resistance as a function of density $n$ and perpendicular magnetic field $B_\perp$ in device M1 and M2 respectively. Apart from the similar dome structures around half-filling as in Fig. 2b-c, there are notably oscillatory features near the boundary between the superconducting phase and the correlated insulator phase. These oscillations can be understood as phase-coherent transport through inhomogeneous regions in the device (see Methods and Extended Data Fig. 1). (c) Differential resistance $dV_{xx}/dI$ versus dc bias current $I$ for different $B_\perp$ values, measured for device M2. (d) $R_{xx}-T$ curves for different $B_\perp$ values, measured for device M1. (e) Perpendicular and parallel critical magnetic field versus temperature for device M1 (50\% normal state resistance). The fitting curves are plotted according to the corresponding formulas in Ginzburg-Landau theory for a 2D superconductor. Measurements in (a-c) are all taken at \SI{70}{\milli\kelvin}.}
\end{figure*}

The application of a perpendicular magnetic field $B_\perp$ to a 2D superconductor creates vortices that introduce dissipation and gradually suppresses superconductivity.\cite{tinkham1996} Figure 3a and 3b show the resistance of device M1 and M2 as functions of density and $B_\perp$. Both devices exhibit a maximum critical field of approximately \SI{70}{\milli\tesla}. The critical field varies strongly with doping density, showing two similar domes on each side of the half-filling state. Near the Mott-like insulating state (\num{-1.47} to \SI{-1.67e12}{\per\centi\meter\squared} for M1, \num{-1.25} to \SI{-1.35e12}{\per\centi\meter\squared} for M2), periodic oscillations of the resistance and critical current as a function of $B_\perp$ appear (see Methods and Extended Data Fig. 1 for detailed analysis). The oscillations seem to originate from phase-coherent transport through arrays of Josephson junctions, similar to SQUID-like superconductor rings around one or more insulating islands. Such junctions regions may be due to slight density inhomogeneities in the devices, such that a few `islands' are doped into the insulating phase while other parts of the device remain superconducting. Apart from these oscillatory behaviors near the boundary of the half-filling insulating state, the critical current and zero resistivity inside the domes are gradually suppressed by $B_\perp$ as shown in Fig. 3c and 3d. Figure 3e shows the critical magnetic field versus temperature for device M1, under perpendicular and parallel field configurations. The temperature dependence of the perpendicular critical field $B_{c\perp}$ fits well to the Ginzburg-Landau (GL) theory $B_{c\perp}=\frac{\Phi_0}{2\pi\xi_\mathrm{GL}^2}\left(1-\frac{T}{T_c}\right)$, where $\Phi_0=h/2e$ is the superconducting flux quantum, and gives the GL superconducting coherence length $\xi_\mathrm{GL}\approx\SI{52}{\nano\meter}$ (at $T=0$). The in-plane critical field dependence, on the other hand, is not explained by the GL theory for thin-film superconductors, due to the atomic thickness of TBG, $d\approx\SI{0.6}{\nano\meter}$, which would imply an in-plane critical field $B_{c\parallel}\ge\SI{36}{\tesla}$ as the temperature approaches zero. \cite{tinkham1996} Instead, we interpret the dependence of $T_c$ on $B_{\parallel}$ as a result of paramagnetic pair-breaking due to the Zeeman energy. The zero-temperature in-plane critical field is extrapolated to be around \SI{1.1}{\tesla}, which is higher than but close to  the estimated value for the Pauli limit $B_P \approx \SI{1.85}{\tesla\per\kelvin} \cdot T_c\approx\SI{0.93}{\tesla}$ based on the BCS gap formula $\Delta\sim1.76k_BT_c$. We note that the superconductor-metal transition in MA-TBG is not sharp, and therefore the extraction of both $B_c$ and $T_c$ has some uncertainty. Qualitatively, the dependence of the in-plane critical field on temperature is $B_{c\parallel} \sim \left(1-T/T_c\right)^{1/2}$ near $T_c$. \cite{klemm1975}The results shown above are consistent with the existence of 2D superconductivity confined in an atomically thin space. As we will show, the coherence length $\xi$ is comparable to the inter-particle spacing and may suggest that the system is driven close to a BCS-BEC crossover.

\section*{Phase Diagram of MA-TBG}

\begin{figure*}
\includegraphics[width=\textwidth]{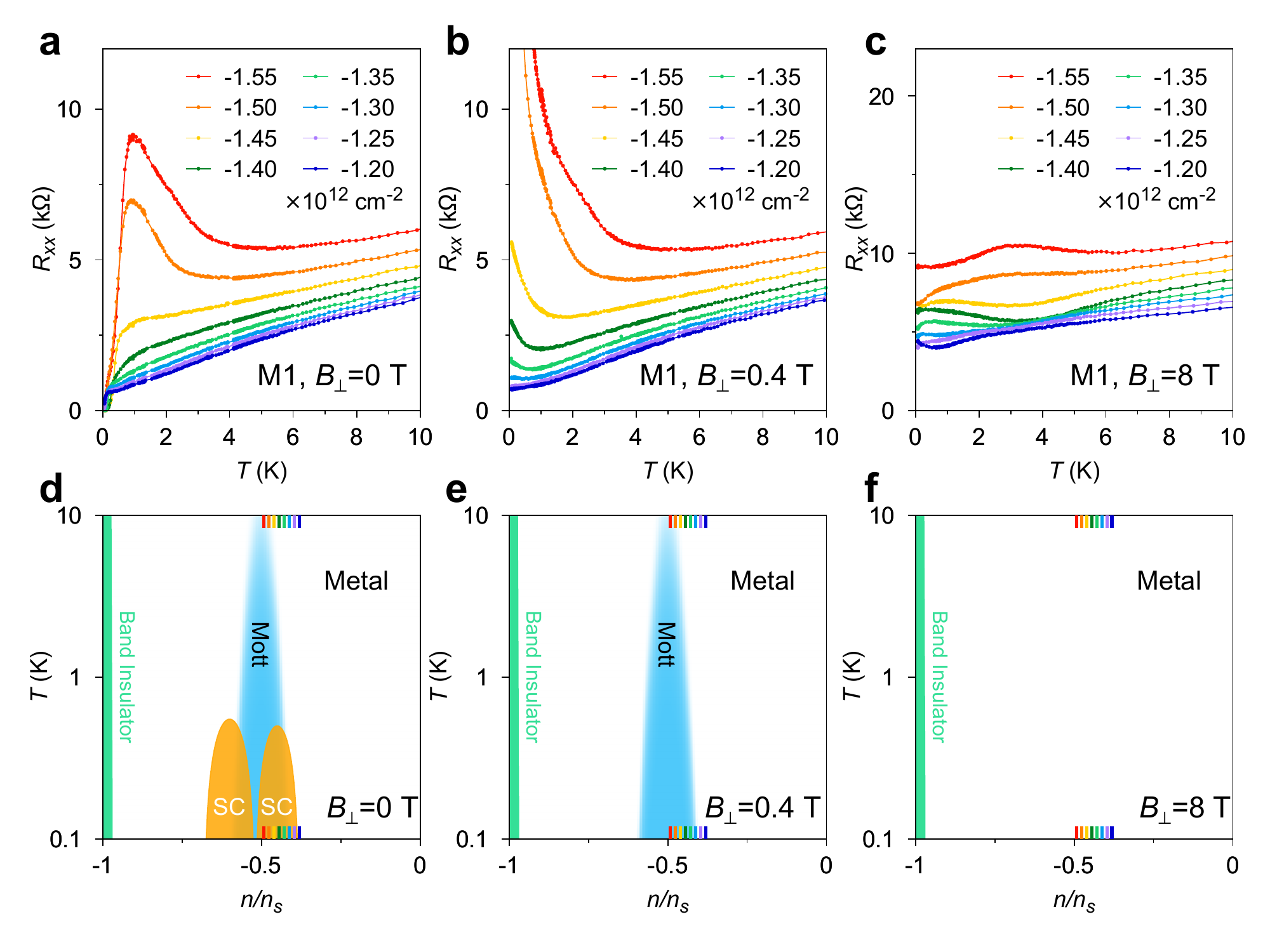}
\caption{\textbf{Temperature-density phase diagrams of MA-TBG at different magnetic fields.} (a-c) $R_{xx}$-$T$ curves for device M1 at different densities measured in $B_\perp=0$, $B_\perp=\SI{0.4}{\tesla}$ and $B_\perp=\SI{8}{\tesla}$. The magnetic field induces a superconductor-insulator-metal transition at the lowest temperature. (d-f) Schematic phase diagrams corresponding to the magnetic fields in (a-c). The horizontal axis is the relative filling $n/n_s$. Short color lines at the top and bottom of the plots denote the densities corresponding to those plotted in (a-c). }
\end{figure*}

MA-TBG provides a rich phase diagram with interplaying correlated insulator phases and superconducting phases, with continuous tunability by temperature, magnetic field, and charge density. Apart from the superconducting domes discussed above, the correlated Mott-like insulator phase at half-filling also assumes a dome shape with a transition to a metallic phase at about 4$\sim$\SI{6}{\kelvin}, and centered around half-filling density. We have previously shown that the MOtt-like insulator phase crosses over to a metallic phase by applying a strong magnetic field around \SI{6}{\tesla} in any direction (perpendicular or parallel to the devices). \cite{cao2018} A plausible explanation is that the many-body charge gap is closed by the Zeeman energy. Figures 4a-4c show the resistance versus temperature data measured in device M1 at zero magnetic field, at $B_\perp=\SI{0.4}{\tesla}$, and at $B_\perp=\SI{8}{\tesla}$, respectively. At zero field, we observe the transition from a metal at high temperatures ($>$\SI{5}{\kelvin}) to a superconductor. Close to half-filling there is an intermediate region with insulating temperature dependence from 1$\sim$\SI{4}{\kelvin} (i.e. above $T_c$), which is identified with the Mott-like insulating phase at half-filling. In a small magnetic field $B_\perp=\SI{0.4}{\tesla}$ above the critical magnetic field, the system remains an insulator down to zero temperature near half-filling, and remains a metal away from the half-filling. Finally, in a strong magnetic field $B_\perp=\SI{8}{\tesla}$, the correlated insulator phase is fully suppressed by the Zeeman effect and the system is metallic everywhere between $n=-n_s$ and the charge neutrality point. Our data shows that a rich phase space of metal-insulator-superconducting physics \cite{goldman2011} is present in MA-TBG. A schematic evolution of the phase diagram as the magnetic field increases is shown in Figs. 4d-f. 

\section*{Quantum Oscillations in the Normal State}

\begin{figure*}
\includegraphics[width=\textwidth]{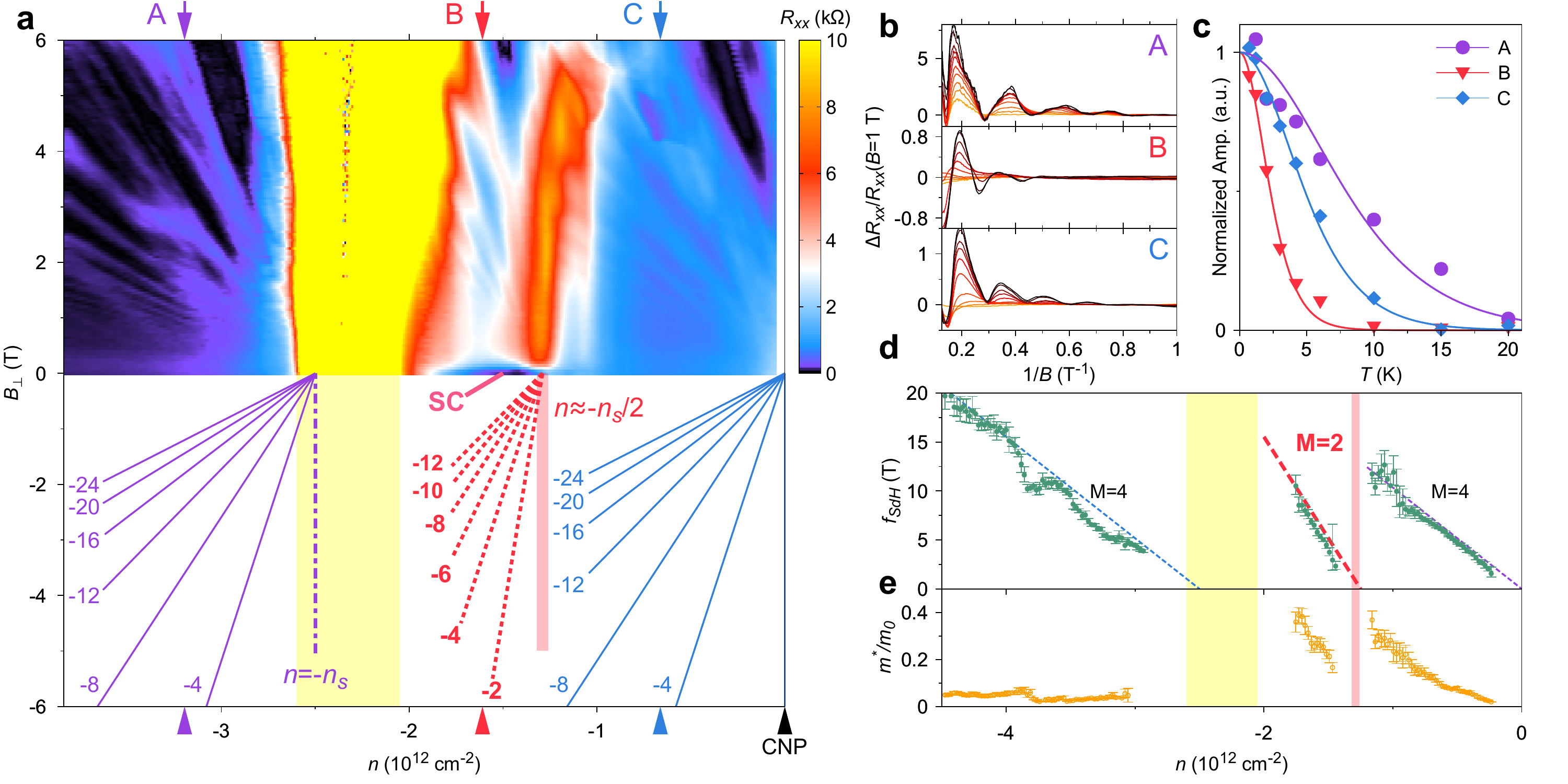}
\caption{\textbf{Quantum oscillations in MA-TBG at high fields.} (a) Magnetoresistance $R_{xx}$ versus density $n$ (hole-doping side with respect to charge neutrality) and $B_\perp$ in device M2. The lower half of the diagram shows the Landau level structure deduced from the oscillations. The blue Landau fan, originating from the charge neutrality point (CNP) on the right, and the purple Landau fan, originating from the superlattice density ($n=-n_s$) on the left, show filling factor sequences $-4, -8, -12, \ldots$ as expected from the single-particle band structure with fourfold spin and valley degeneracies. The extra red fan in the middle, that originates from $-n_s/2$, however, has a peculiar sequence of $-2, -4, -6, \ldots$ and is not expected from the single particle band structure. (b) Temperature-dependent quantum oscillation traces at charge densities labeled by A, B, and C in (a). From black to orange ordered by brightness, the temperatures are \SI{0.7}{\kelvin}, \SI{1.2}{\kelvin}, \SI{2.0}{\kelvin}, \SI{3.0}{\kelvin}, \SI{4.2}{\kelvin}, \SI{6}{\kelvin}, \SI{10}{\kelvin}, \SI{15}{\kelvin}, \SI{20}{\kelvin} and \SI{30}{\kelvin}. (c) Lifshitz-Kosevich fit for the normalized amplitudes of the oscillations shown in (b). (d) and (e) are extracted SdH oscillation frequencies and effective masses as a function of charge density $n$.}
\end{figure*}

We have studied quantum oscillations in the entire accessible density range, including in the vicinity of the correlated insulating state where superconductivity occurs. Figures 5a-b show the Shubnikov-de Haas (SdH) oscillations in longitudinal resistance as a function of carrier density on the overall hole-doping region ($E_{F}<0$) for device M2. The Landau levels in a TBG superlattice typically follow the equation $n/n_s = N\phi/\phi_0 + s$, where $\phi=B_\perp A$ is the magnetic flux penetrating each unit cell and $\phi_0=h/e$ is the (non-superconducting) flux quantum. $N=\pm1,2,3,\ldots$ is the Landau level index. $s=0$ denotes the Landau fan that emanates from the Dirac point, and $s=\pm1$ denote the Landau fans that result from electron-like or hole-like quasiparticles near the band edges of the single-particle superlattice bands in the MBZ, which emanate from $\pm n_s$. These Landau levels also exhibit a fourfold degeneracy due to spins and valleys, and thus the filling factor sequence $4, 8, 12, \ldots$. 

Surprisingly, in addition to these expected Landau fans, we observe an additional Landau fan that emanates from the correlated insulating state at $-n_s/2$. This Landau fan has $N=-1/2,-1,-3/2,-2,\ldots$ (\emph{i.e.} filling factors $-2, -4, -6, -8, \ldots$) and $s=-1/2$. The superconducting dome is distinguishable in Fig. 5a directly beneath this Landau fan, being very close to zero field and next to the correlated insulating region. Unlike commonly observed broken-symmetry states that split from a single degenerate Landau level into multiple levels, the halved filling factors appear to be intrinsic to the fan and hold down to the lowest magnetic field where the oscillations are still visible. Fractional values for $s$ have been reported in graphene superlattices due to Hofstadter's butterfly, which typically occurs in much stronger magnetic fields ($>\SI{10}{\tesla}$) and only become obvious at the intersection of Landau levels with different integer $s$.\cite{hunt2013,ponomarenko2013,dean2013} Therefore, Hofstadter's butterfly physics cannot explain the extra stand-alone fan observed here, which appears at fields as low as \SI{1}{\tesla}. Furthermore, the halving of the filling factors and $s$ is unlikely to be explained in a single-particle picture of unit cell doubling due to strain or formation of a charge density wave, in which cases either spin or valley degeneracy must be broken \emph{a priori}. We have observed the same Landau level sequence in two other MA-TBG devices, so it is robust against small variations in twist angle and consistent across samples (see Methods and Extended Data Fig. 2).

To further study the non-trivial origin of the Landau fan near half-filling, we measured the effective mass from the temperature-dependent quantum oscillation amplitude according to the Lifshitz-Kosevich (LK) formula (see Methods for definition). Figures 5b-c show the oscillations and oscillation amplitudes at three different densities indicated by arrows in Fig. 5a. Figures 5d-e show the oscillation frequency $f_{SdH}$ and effective mass extracted by fitting the oscillation amplitudes to the LK formula. The dependence of $f_{SdH}$ on charge density $n$ provides another perspective on the exotic oscillations, since the value of $M=\phi_0\Delta n/\Delta f_{SdH}$ extracted from the slope directly provides the number of degenerate Fermi pockets. The experimental data clearly fits to $M=4$ near the charge neutrality point and for densities beyond the superlattice gap, while $M=2$ for the quantum oscillations starting near the correlated insulator state and right above the superconducting dome. The effective mass of the anomalous oscillations is \SIrange{0.2}{0.4}{}$m_0$ where $m_0$ is the bare electron mass, significantly higher than the mass near charge neutrality ($\sim0.1m_0$) and beyond the superlattice gap ($\sim0.05m_0$), at the same relative density $\Delta n$ from where $f_{SdH}$ goes to zero.

The quantum oscillations above the superconducting dome clearly indicate the existence of small Fermi surfaces that originate from the correlated insulating state with an area proportional to $n^\prime=|n|-n_s/2$, instead of a large Fermi surface that correspond to the density $|n|$ itself. The Hall measurements which we show in the Extended Data Fig. 3 also support this conclusion. Notably, similar small Fermi pockets that do not correspond to any pockets in the single-particle Fermi surface have been observed in under-doped cuprates, \cite{yelland2008, bangura2008, jaudet2008} although the origin of these peculiar small Fermi pockets is still debated. Among the possibilities, the observed small Fermi surface can be the Fermi surface of quasiparticles that are created by doping a Mott insulator. \cite{lee2006, kaul2008} The halved degeneracy, on the other hand, might be related to spin-charge separation as predicted in a doped Mott insulator. \cite{kaul2008} However, more experimental and theoretical work is needed to clarify the origin of these intriguing quantum oscillations.

\section*{Discussion}

\begin{figure*}
\includegraphics[width=\textwidth]{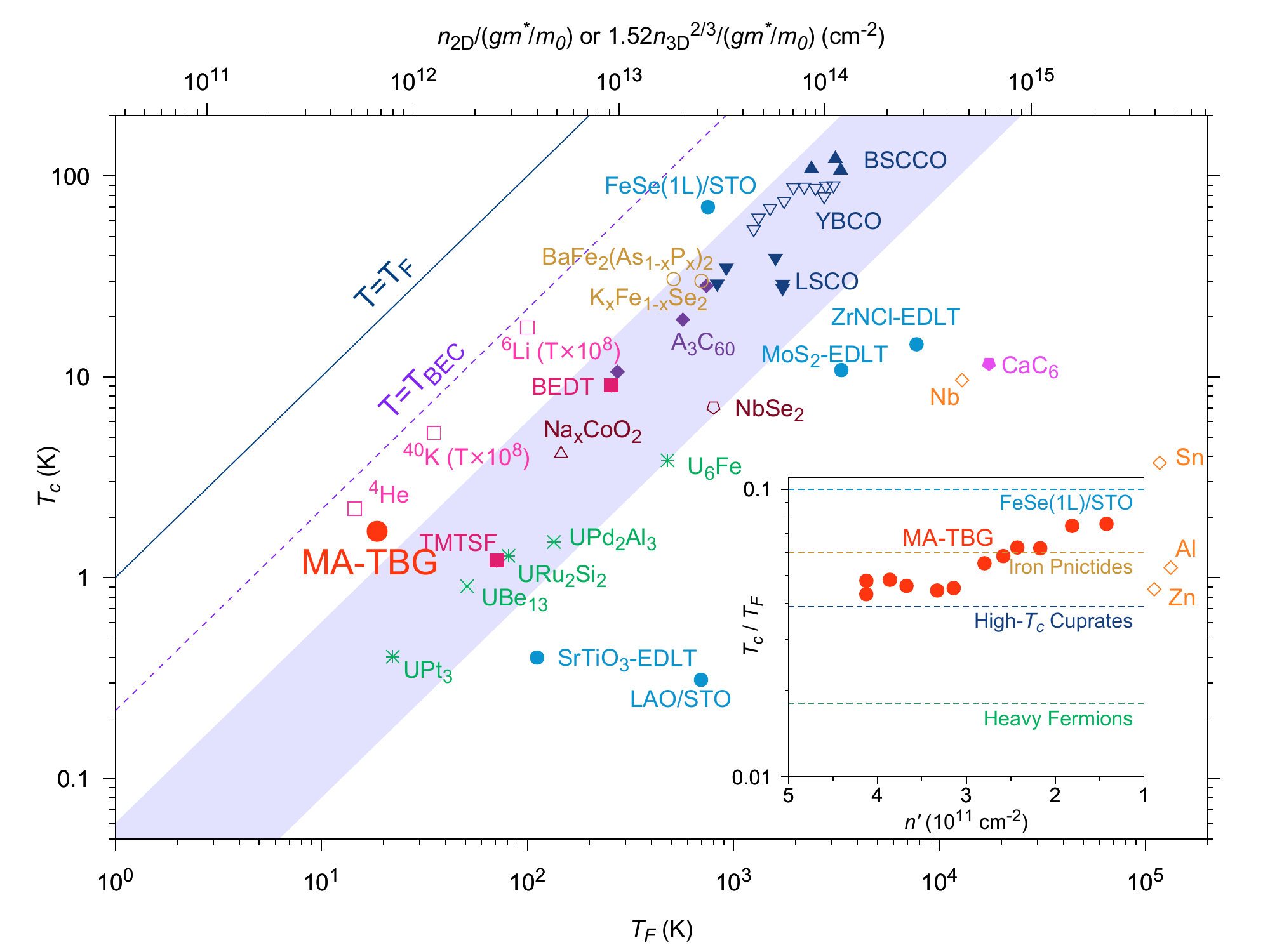}
\caption{\textbf{Superconductivity in the strong-coupling limit.} Log-log plot of $T_c$ versus Fermi temperature $T_F$ for various superconductors.\cite{uemura2004} The top axis is the corresponding 2D carrier density $n_{2D}$ for 2D materials or $n_{3D}^{2/3}$ for 3D materials, normalized with effective mass $m^*$ and Fermi surface degeneracy $g$ (and a constant factor for 3D density). 2D superconductors in this plot are represented by solid circular dots. $T_\mathrm{BEC}=1.04 \hbar^2 n_{3D}/m^*$ for a 3D bosonic gas is plotted for comparison. Bose-Einstein condensation temperatures in \textsuperscript{4}He, and paired fermionic \textsuperscript{40}K and  \textsuperscript{6}Li (both axes multiplied by $10^8$ for the latter two) are shown as empty squares. \cite{uemura2004,ku2012} The point for MA-TBG is calculated from the 2D density and effective mass obtained from quantum oscillations (Fig. 5d-e) at the optimal doping, $n_{2D}=\SI{0.15e12}{\per\centi\meter\squared}$ and $m^*=0.2m_0$, and using $g=1$ accounting for the halved degeneracy. Inset shows the variation of $T_c/T_F$ as a function of charge doping for MA-TBG. Data for other materials are either adapted from Uemura, et. al. \cite{uemura2004} or extracted from the literature. \cite{qian2011,hashimoto2012,saito2015,ye2012,caviglia2008,ueno2008,weller2005,valla2009,peelaers2012,mccollam2014}.
}
\end{figure*}

The appearance of both superconductor and correlated insulator phases in the flat bands of MA-TBG at such a small charge density is beyond weak-coupling BCS theory. The carrier density that is responsible for $T_c=\SI{1.7}{\kelvin}$ is extremely small according to the quantum oscillation measurements, merely $n^\prime=\SI{1.5e11}{\per\centi\meter\squared}$ at optimal doping. To place this in the context of other superconductors, we plot $T_c$ against $T_F$ in a log-log plot\cite{uemura2004} for various materials in Fig. 6, where $T_F$ is the Fermi temperature and proportional to the 2D carrier density $n_{2D}$, which the quantum oscillations data show to be $n^\prime$ for the superconducting dome region of MA-TBG. Most unconventional superconductors have $T_c/T_F$ values of 0.01$\sim$0.05, while all the conventional BCS superconductors lie on the far right in the plot, with much smaller ratios. MA-TBG is located above the trend line on which most cuprates, heavy fermion and organic superconductors lie, approaching the recently observed exotic FeSe monolayer on SrTiO\textsubscript{3} (see Fig. 6 inset), strongly suggesting that the superconductivity originates from electron correlations instead of weak electron-phonon coupling.  One other frequently compared temperature is the Bose-Einstein condensation (BEC) temperature for a 3D boson gas $T_\mathrm{BEC}$, assuming that all particles in the occupied Fermi sea pair up and condense. Cuprates and other unconventional superconductors typically have $T_c/T_\mathrm{BEC}$ ratios on the order of 0.1$\sim$0.2. The $T_c/T_\mathrm{BEC}$ for MA-TBG is estimated to be up to 0.37, indicating very strong electron-electron interactions and possibly close proximity to the BEC-BCS crossover. This is in agreement with the fact that the coherence length in MA-BLG, of order $\xi\sim\SI{50}{\nano\meter}$ at optimal doping, is on the same order of magnitude as the average inter-particle distance $\sim n^{\prime-\frac{1}{2}}\approx\SI{26}{\nano\meter}$.

In summary, the realization of unconventional superconductivity in a graphene superlattice establishes MA-TBG as a relatively simple, clean, accessible, and, more importantly, highly tunable platform for studying correlated electron physics. The interactions in MA-TBG can possibly be further fine-tuned by the twist angle \cite{cao2018} and by the application of perpendicular electrical fields by means of differential gating \cite{gonzalez2017}. Moreover, $T_c$ can possibly be further enhanced by applying pressure to the graphene superlattice in order to increase the interlayer hybridization \cite{yankowitz2017} or by coupling different MA-TBG structures to induce Jospehson coupling in the vertical direction. Moreover, similar magic angle superlattices and flat-band electronic structures can be realized with other 2D materials or lattices to investigate strongly-correlated systems with different properties. Finally, we comment that despite the several apparent similarities to cuprates phenomenology, there are key differences with our MA-TBG superlattice realization: the first is that the valley degree of freedom in the underlying graphene lattices leads to an extra degeneracy, resulting in two carriers per super-lattice unit cell at half-filling in the parent correlated insulator state, though higher quality devices and fine tuning may lead to superconductivity also near the one and three carriers per unit cell regions. The second key difference is that in MA-TBG the underlying superlattice is triangular, which should have a fundamental influence on the type of spin-singlet ground state it can possibly host, due to magnetic frustration. The lattice symmetry should also impose certain limitations on the possible superconducting pairing symmetry in MA-TBG, which requires further experiments such as tunneling and Josephson hetero-junctions to confirm.\cite{tsuei2000} Various pairing symmetries including $d+id^\prime$-wave, $p_x+ip_y$-wave and spin-triplet $s$-wave have been theoretically predicted in the hypothetical superconductivity of monolayer or few-layer graphene. \cite{nandkishore2012, uchoa2007, hosseini2012}Among these possibilities, if the mechanism for superconductivity in MA-TBG is indeed related to the correlated half-filling insulating state, as it is the case in $d_{x^2-y^2}$-wave cuprates, the pairing symmetry might possibly be chiral $d+id^\prime$ wave in order to satisfy the underlying triangular lattice symmetry of the superlattice. We hope that further experimental and theoretical work on MA-TBG and related MA superlattices will stimulate new insight on the key ingredients that govern unconventional superconductivity, as well as bring us closer to the realization of tunable quantum spin liquids. \cite{balents2010}

\section*{Acknowledgements}
We acknowledge helpful discussions with R. Ashoori, S. Carr, R. Comin, L. Fu, P. A. Lee, L. Levitov, K. Rajagopal, S. Todadri, A. Vishwanath, and M. Zwierlein. This work has been primarily supported by the Gordon and Betty Moore Foundation's EPiQS Initiative through Grant GBMF4541 and the STC Center for Integrated Quantum Materials (NSF Grant No. DMR-1231319) for device fabrication, transport measurements, and data analysis (Y.C., V.F., P.J.-H.), as well as theoretical calculations (S.F.). K.W. and T.T. acknowledge support from the Elemental Strategy Initiative conducted by the MEXT, Japan and JSPS KAKENHI Grant Numbers JP15K21722 and JP25106006. This work made use of the Materials Research Science and Engineering Center Shared Experimental Facilities supported by the National Science Foundation (DMR-0819762) and of Harvard's Center for Nanoscale Systems, supported by the NSF (ECS-0335765). E.K. acknowledges additional support by ARO MURI Award W911NF-14-0247.

\clearpage

\onecolumngrid
\section*{Methods}

\renewcommand{\figurename}{Extended Data Fig.}
\setcounter{figure}{0}

\subsection*{Sample Preparation}
The devices are fabricated using a modified dry-transfer technique detailed in previous work \cite{cao2016,cao2018,kim2016}. Monolayer graphene and hexagonal boron nitride (\SIrange{10}{30}{\nano\meter} thick) are exfoliated on SiO\textsubscript{2}/Si chips and high-quality flakes are picked using optical microscopy and atomic force microscopy. We use Poly (Bisphenol A carbonate) (PC)/Polydimethylsiloxane (PDMS) stack on a glass slide mounted on a home-made micro-positioning stage to first pick up an h-BN flake at \SI{90}{\celsius}, and then uses the van der Waals force between h-BN and graphene to tear a graphene flake at room temperature. The separated graphene pieces are manually rotated by a twist angle $\theta$ about \SIrange{1.2}{1.3}{\degree} and stacked together again, resulting in a precisely controlled TBG structure. The stack is encapsulated with another h-BN flake on the bottom and released onto a metal gate at \SI{160}{\celsius}. We perform no heat annealing after this step since we find that TwBLG tend to relax to Bernal-stacked bilayer graphene at high temperatures. The final device geometry is defined by electron-beam lithography and reactive ion etching. Electrical connections are made to the TBG by Cr/Au edge-contacted leads \cite{wang2013}.

\subsection*{Measurements}

Transport measurements are performed in a dilution refrigerator with a base temperature of $\sim$\SI{70}{\milli\kelvin} except for the temperature-dependent quantum oscillations which are measured in a He-3 fridge. 
We use standard low frequency lock-in techniques with excitation frequency of \SIrange{5}{10}{\hertz} and excitation current of \SIrange{0.4}{5}{\nano\ampere}. The current flowing through the sample is amplified by a current pre-amplifier and measured by the lock-in amplifier. Four-probe voltage is amplified by a voltage pre-amplifier at $\times1000$ and measured by another lock-in amplifier. 

The twist angle of the devices is determined from the transport measurements at low temperatures, which are detailed in the Methods section of our previous work \cite{cao2018}. Briefly, a rough estimate of the twist angle can be given by the carrier density of the superlattice gaps at $\pm n_s$ which exhibit as strongly insulating states. To refine this estimate, the Landau levels that appear at high magnetic fields are fitted to the Wannier diagram, which gives the twist angle with uncertainty of about 0.01$\sim$\SI{0.02}{\degree}. 

\subsection*{Extraction of quantum oscillation frequency and effective mass}

The effective mass in device M2 is extracted using the standard Lifshitz-Kosevich formula, which relates the temperature-dependence of resistance change $\Delta R_{xx}(T)$ to the cyclotron mass $m^*$ (at a given magnetic field $B_\perp$).
\begin{equation}
\label{eq:lk}\Delta R_{xx}(T) \propto \frac{\chi}{\sinh(\chi)}, \quad \chi=\frac{2\pi^2 k_B T m^*}{\hbar e B_\perp}.
\end{equation}
For each gate voltage (charge density), we measure the $R_{xx}$--$B_\perp$ curves at different temperatures, normalize them by their low-field values, and subtract a common polynomial background in $B_\perp$. Examples of the curves are shown in Fig. 5b. The oscillation frequencies shown in Fig. 5d are extracted from these curves plotted versus $1/B_\perp$. From the temperature-dependent amplitude of the most prominent peak, we extract $m^*$ using the above formula (Fig. 5e). The errorbars in Fig. 5d and 5e represent 90\% confidence intervals of the fitting.

\subsection*{Commensuration and twist angle}
Mathematically, in a twisted moir\'{e} system, the lattice is strictly periodic only when the twist angle satisfies a specific relation such that lattice registration order is perfectly recovered in a finite distance. These special cases are termed `commensurate' structures. One important parameter in the commensurate TBG structures is $r$, which can be intuitively understood as the number of `apparent' moir\'{e} pattern wavelengths it takes to fully recover the lattice periodicity. \cite{shallcross2010, santos2012} The simplest commensurate structures with $r=1$ are called `minimal' structures. These structures have exactly one moir\'{e} spot per unit cell. In TBG, apart from the minimal structures which only occur at discrete angles, there are other commensurate structures that are arbitrarily close to any given angle $\theta$ with large $r$. At small twist angles however, it has been shown that the evolution of the band structure of TBG can be effectively viewed as semi-continuous, \emph{i.e.} an infinitesimal change of twist angle does not have any significant effect in the band structure even though the lattice can be in a different `family' of commensurate structures (different $r$). \cite{santos2012} In other words, the TBG system can be well-approximated by a continuum model, which was originally proposed by Bistritzer \emph{et. al.} \cite{bistritzer2011}, and the physics in minimal structures is representative of all nearby commensurate structures. In our experiments, we do not expect the lattice to be in absolutely perfect commensuration due to disorder and intrinsic randomness due to the fabrication process, so we think that the continuum model can faithfully represent the realistic TBG system where any commensuration effect has been smoothed out.

We deduce the size of the moir\'{e} unit cell and twist angle based on the density of the superlattice gaps $\pm n_s$ ($\pm4$e\textsuperscript{-}/moir\'{e}), and then cross-check the twist angle with the Landau levels observed at high magnetic fields. $\pm n_s$ are the only multiple of $n_s$ that correspond to Fermi energies located within single-particle band gaps and therefore exhibit strong insulating behavior. For twist angles above about \SI{0.9}{\degree} to \SI{1}{\degree}, the band structure at energies higher than these gaps is very complex (with strongly overlapping bands) and no single particle gap at $\pm2n_s,\pm3n_s,\ldots$ appears. \cite{bistritzer2011, santos2012, nam2017, morell2010, moon2012} The experimentally measured values for the single-particle insulating gaps we observe are in the 30$\sim$\SI{60}{\milli\electronvolt} range.\cite{cao2016,cao2018} Below about \SI{0.9}{\degree} to \SI{1}{\degree}, however, the superlattice gaps at $\pm n_s$ close and there is no single-particle gap \emph{at any energy} in the system. \cite{nam2017, kim2017} In this regime, there are Dirac-like bands crossing at $\pm2n_s$ which might be responsible for the resistance peaks observed in devices with very small twist angles, though possible interaction effects may enhance these \cite{kim2017}. However these states observed in very low-twist devices are clearly different from the strong insulating gaps observed in our present work and previous work.\cite{cao2018} We emphasize that there is a drastic change in the band structure at about \SI{0.9}{\degree} to \SI{1}{\degree} (depending on the parameters of the model being used), leading to a transition from single particle gaps at $\pm n_s$  to resistive states at $\pm 2n_s$. This crossover can be clearly observed in the included supplementary video, where we show an evolution of the band structure of TBG from $\theta=\SI{3}{\degree}$ to $\theta=\SI{0.8}{\degree}$. The data in the video is calculated using the continuum model described in Bistritzer \emph{et. al.} \cite{bistritzer2011}

\subsection*{Possible effects due to finite electrical fields}
It has been shown that by applying a perpendicular electrical field to Bernal-stacked bilayer graphene, topological states can exist on the AB/BA stacking boundaries while the bulk of the AB and BA regions remain gapped. \cite{zhang2013, vaezi2013, ju2015} In small angle TBG, a similar effect can affect the band structure because the AA-stacked regions in the moir\'{e} pattern are interconnected by these AB/BA staciking boundaries. This effect has been recently observed in scanning tunneling experiments on ultra-small twist angle samples.\cite{huang2018}

One may ask then how will the flat bands in MA-TBG be affected by the network of topological boundaries when a residual electrical field is present. Theoretical work on $\theta=\SI{1.5}{\degree}$ TBG has shown that when an inter-layer potential difference of $\Delta V=\SI{300}{\milli\electronvolt}$ is applied, the low-energy superlattice bands become even flatter and the electronic states become more localized. \cite{gonzalez2017} Therefore, there is good reason to believe that the flat-band physics presented in the present article holds even when a perpendicular electric field is present, since the electric field is likely going to render the band structure even more localized and correlated as one gets closer to the magic angle. In our experiments, in any case, we estimate that the potential difference between the two layers induced by our gate voltage is at the very most on the order of \SI{50}{\milli\volt}, and likely much smaller due to screening. Any possible effects of the residual electric field should be minimal. 

\subsection*{Phase-coherent transport behavior in superconducting MA-TBG}

In Fig. 3a-b of the main text, we observe oscillatory behavior in the measured longitudinal resistance $R_{xx}$ as a function of perpendicular magnetic field $B_\perp$, when the charge density is close to the boundary between the half-filling insulating state and the superconducting states. The oscillations are most clearly seen for $n=-1.70\sim$\SI{-1.60e12}{\per\centi\meter\squared} and $n=-1.50\sim$\SI{-1.47e12}{\per\centi\meter\squared} in device M1.

\begin{figure}
\includegraphics[width=0.8\textwidth]{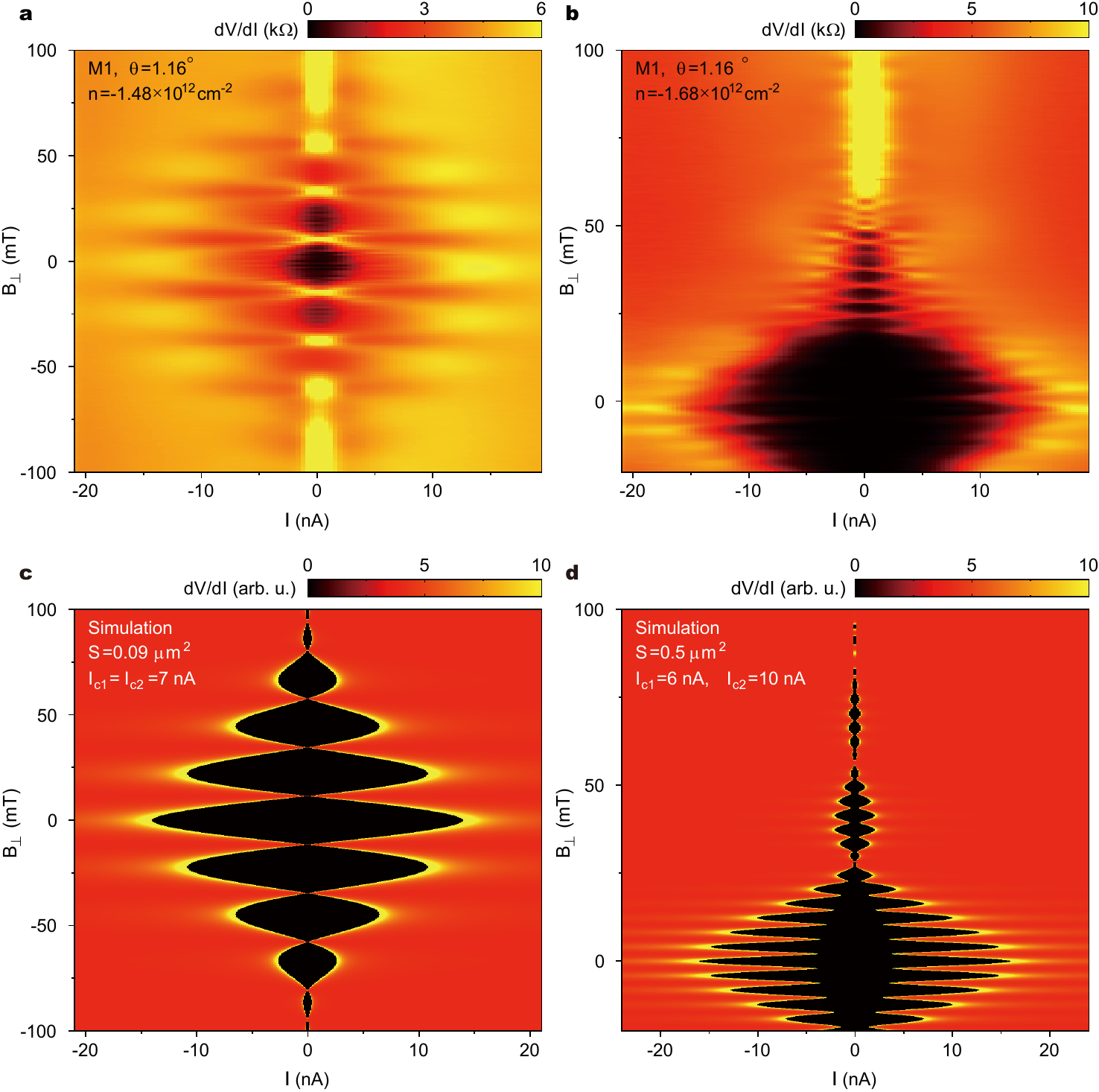}
\caption{\textbf{Evidence of phase-coherent transport in the superconducting MA-TBG. } (a-b) Differential resistance versus bias current $I$ and perpendicular field $B_\perp$, at two different charge densities $n$ corresponding to Fig. 3a in the main text. Periodic oscillations are observed in the critical current. (c-d) Simulations intended to qualitatively reproduce the behavior observed in (a-b).}
\end{figure}

Extended Data Fig. 1a-b show the differential resistance $\mathrm{d}V_{xx}/\mathrm{d}I$ versus bias current $I$ and perpendicular magnetic field $B_\perp$. At zero bias current, the oscillation of the differential resistance with $B_\perp$ in Extended Data Fig. 1a and 1b correspond to line cuts in Fig. 3a at densities $n=\SI{-1.48e12}{\per\centi\meter\squared}$ and \SI{-1.68e12}{\per\centi\meter\squared} respectively. The critical current, above which the superconductor turns normal, also oscillates with $B_\perp$ at the same frequency as can be visualized by the bright peaks in Extended Data Fig. 1a-b. The oscillation period is $\Delta B=\SI{22.5}{\milli\tesla}$ in Extended Data Fig. 1a and about $\Delta B=\SI{4}{\milli\tesla}$ in Extended Data Fig. 1b.

The fact that the critical current is maximum at zero $B_\perp$ and oscillates at periodic intervals of the magnetic field suggests the existence of Josephson junction arrays, in the simplest case a SQUID(Superconducting QUantum Interference Device)-like superconducting loop, around a normal or insulating island\cite{tinkham1996}. It is unclear whether this inhomogeneous behaviour is a result of sample disorder or a coexistence of two different phases (\emph{e.g.} the superconducting phase and the correlated insulator phase). Given the 2D nature of our devices, the detailed current distribution in the device cannot be uniquely determined at this moment by transport measurements, but from the oscillation period one can deduce the effective loop area of the SQUID approximately using $S=\Phi_0/\Delta B$ where $\Phi_0=h/2e$ is the superconducting quantum flux. (Note the difference between $\phi_0=h/e$ for quantum Hall effect and $\Phi_0=h/2e$ for superconductivity.) For the experimental data in Extended Data Fig. 1a-b, we obtain areas $S$ of \SI{0.09}{\micro\meter\squared} and \SI{0.5}{\micro\meter\squared} respectively. By comparison, the total device area between the voltage probes is approximately \SI{1}{\micro\meter\squared}.

Using a simple model of a SQUID with a phenomenological decay of the oscillation amplitude at higher magnetic fields, we can attempt to qualitatively reproduce the observed oscillations by numerical simulations. Extended Data Fig. 1c shows the simulated $I$--$B_\perp$ map of the differential resistance for a SQUID with area $S=\SI{0.09}{\micro\meter\squared}$, with the same critical current $I_{c1}=I_{c2}=\SI{7}{\nano\ampere}$ in the two branches, corresponding to experimental data in Extended Data Fig. 1a. Extended Data Fig. 1d shows the simulation for an asymmetric SQUID, of which the area $S=\SI{0.5}{\micro\meter\squared}$ and the critical currents are $I_{c1}=\SI{6}{\nano\ampere}$ and $I_{c2}=\SI{10}{\nano\ampere}$ for the two branches respectively, which accounts for the partial cancellation of the critical current at low fields (\emph{i.e.} the total critical current does not reach zero in an oscillation) as seen in Extended Data Fig. 1b. These simulations are meant to provide a qualitative perspective on these oscillatory phenomena, as the actual supercurrent distribution is likely significantly more complex and will need to be established via magnetic imaging techniques. However, our data indicate that the superconducting behaviour we observe is indeed a phase coherent phenomenon. Although we did not deliberately fabricate SQUID devices using MA-TBG, these periodic oscillations of the critical current in $B_\perp$ are very likely a result of phase-coherent transport (Josephson effect) through a superconductor with insulating puddles and provide further confirm the existence of superconductivity in MA-TBG. 

We also note that, \emph{induced} superconductivity in graphene and graphene-based systems by proximity effect to another superconductor has been demonstrated in graphene since 2007, and graphene-based Josephson junctions continue to be 
explored extensively. \cite{heersche2007, calado2015, bretheau2017} Recently, superconductivity in graphene induced by proximity to a high-$T_c$ superconductor has been reported and indications of induced  unconventional pairing have been observed. \cite{bernardo2017, perconte2018}

\section*{Supplemental quantum oscillation data and low-field Hall effect}

\begin{figure}
\includegraphics[width=\textwidth]{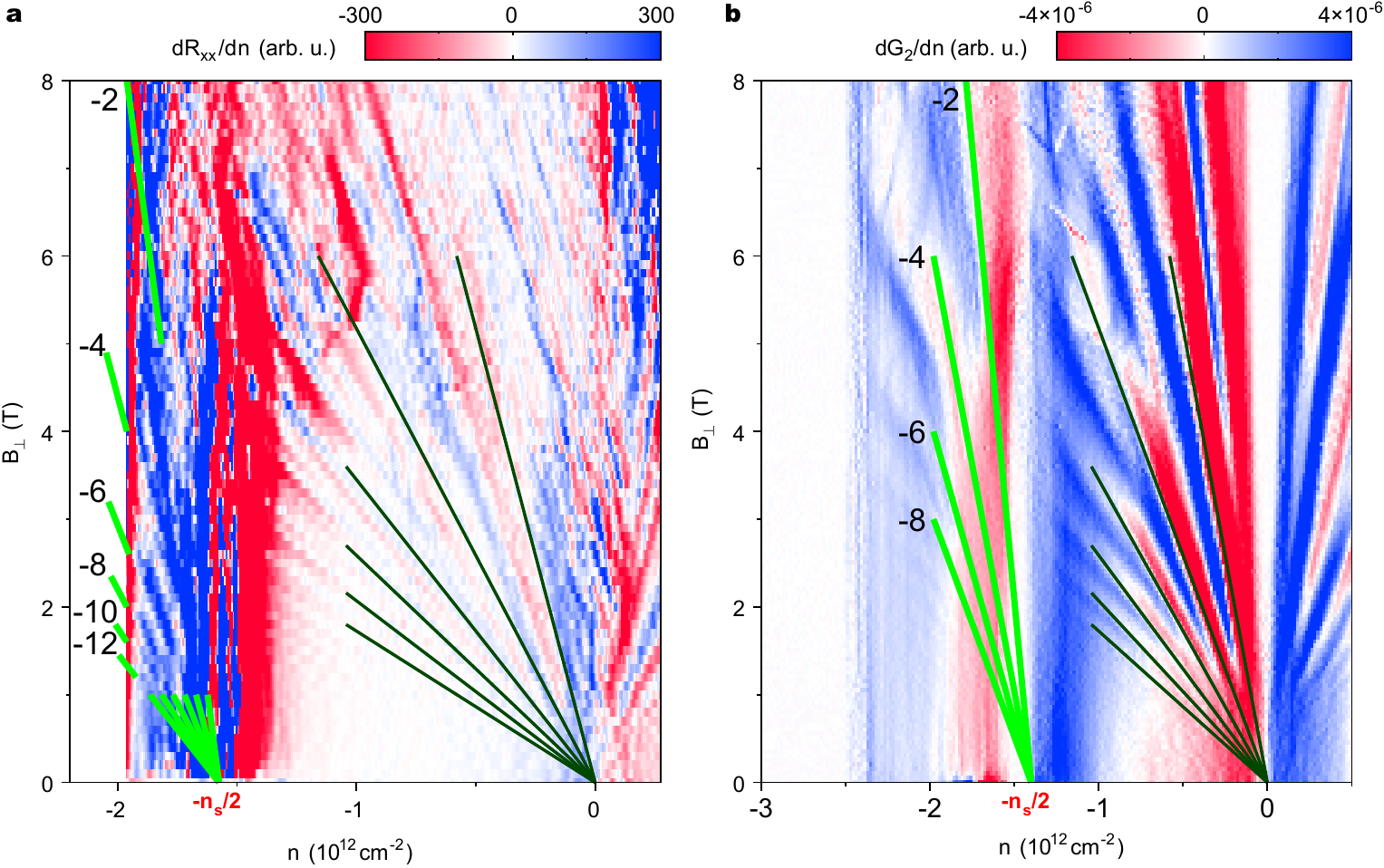}
\caption{\textbf{Supplementary quantum oscillation data.} Quantum oscillations in (a) device M1 ($\theta=\SI{1.16}{\degree}$, data shown for $R_{xx}$) and (b) D1 ($\theta=\SI{1.08}{\degree}$, data shown for two-probe conductance $G_2$). The data is taken the first derivative with respect to the gate-defined charge density $n$ to enhance the color contrast. Both devices exhibit a Landau fan emerging from the half-filling state $-n_s/2$ and have a Landau level sequence of $-2, -4, -6, -8, \ldots$, consistent with the results shown in Fig. 5 of the main text. By comparison, the Landau fan that start from the charge neutrality have a sequence of $-4, -8, -12, \ldots$.}
\end{figure}

Extended Data Figure 2 shows magneto-transport data in device M1 and another magic-angle device D1. Both devices show evidence for the existence of an extra Landau fan with a degeneracy of $M=2$ that emerges from the half-filling insulating states. We note that all magic-angle devices that we have measured so far display quantum oscillations corresponding to emergent quasiparticles only on \emph{one side} of the half-filling states, the one that is away from the charge neutrality point, \emph{i.e.} $n\lessapprox -n_s/2$ for $E_{F}<0$ and $n\gtrapprox n_s/2$ for $E_{F}>0$ (see our previous work for $E_{F}>0$ data\cite{cao2018}), but not observed at $n\gtrapprox -n_s/2$ or $n\lessapprox n_s/2$. The Hall measurements shown below also exhibit a similar asymmetry around the half-filling state. This universal asymmetric behaviour, regardless of the twist angle variation, might be explained if the effective mass of the quasiparticles on the side closer to the charge neutrality is much higher and therefore has a much lower mobility, so that the quantum oscillations cannot be observed and their contribution to the Hall effect becomes negligible. Further theoretical work can possibly shed more light onto the true nature of the many-body energy gap and the related quasiparticles. 

\begin{figure}
\includegraphics[width=0.9\textwidth]{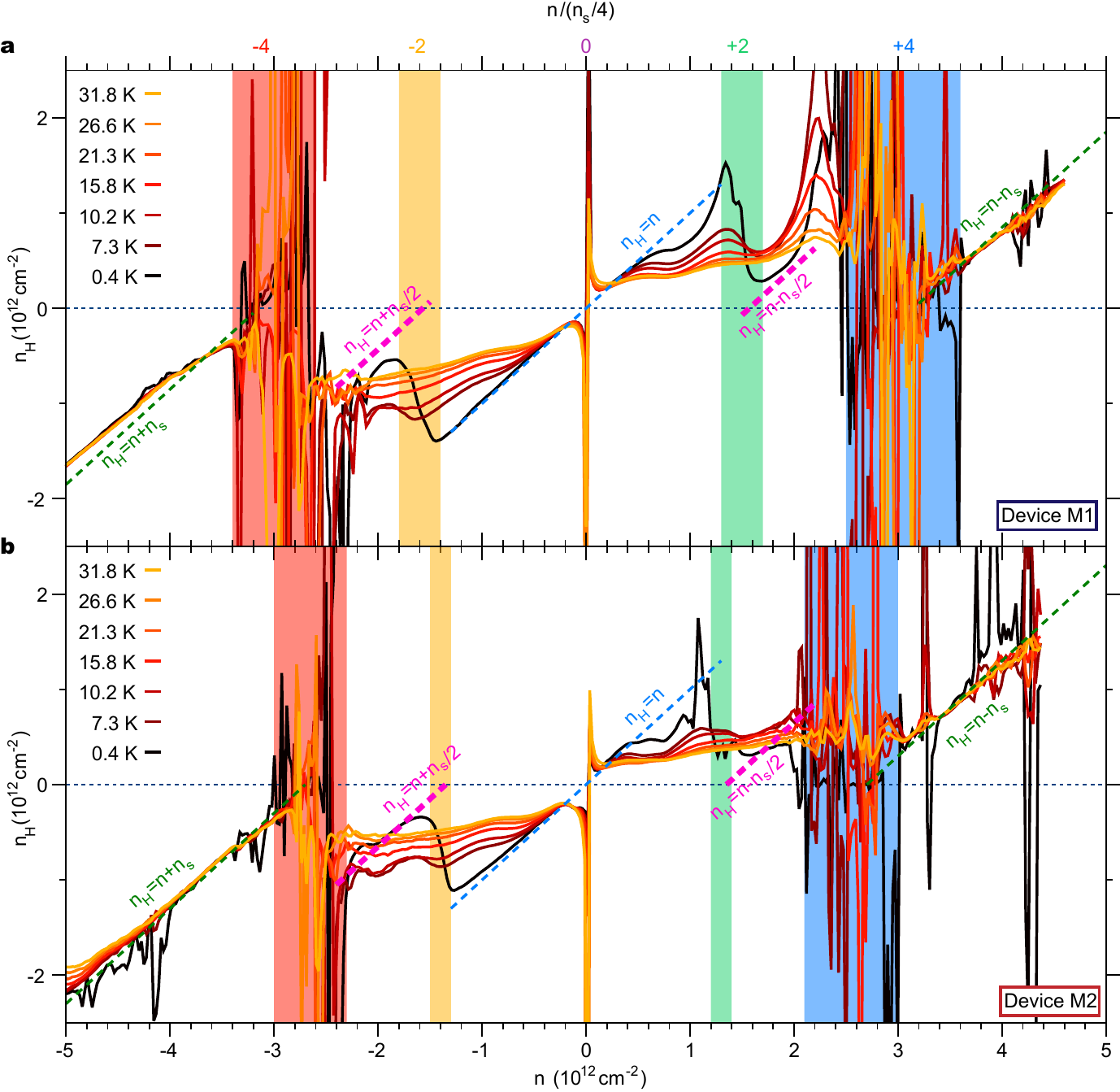}
\caption{\textbf{Low-field Hall effect in MA-TBG. } Low-field Hall effect for device M1 and M2. The Hall density $n_H=-\frac{1}{e}\left(\frac{\mathrm{d}R_{xy}}{\mathrm{d}B_\perp}\right)^{-1}_{B_\perp=0}$ is plotted as a function of total charge density induced by the gate, $n$, measured at temperatures from \SI{0.4}{\kelvin} to \SI{31.8}{\kelvin}. Colored vertical bars correspond to densities $-n_s$, $-n_s/2$, $n_s/2$, and $n_s$ for the two samples respectively.}
\end{figure}

In the main text, we have determined the Fermi surface area in the MA-TBG devices using the Shubnikov-de Hass (SdH) oscillation frequency in a magnetic field (Fig. 5). We find that novel oscillations emerge from the correlated insulating state at half-filling $n=-n_s/2$, and the oscillation frequency indicates small Fermi pockets associated with a shifted density $n^\prime=|n|-n_s/2$.

Extended Data Figure 3 shows another measurement of the transport carrier density via the low-field Hall effect measured up to \SI{+-1}{\tesla}. The measured Hall density, given by $n_H=-\frac{1}{e}\left(\frac{\mathrm{d}R_{xy}}{\mathrm{d}B_\perp}\right)^{-1}_{B_\perp=0}$, provides an independent measurement of the carrier density in the system. In both devices at a temperature of \SI{0.4}{\kelvin}, we observe that while the Hall density closely follows the gate induced density, \emph{i.e.} $n_H=n$, near charge neutrality and up to the half-filling insulating states at $|n|=n_s/2$, it `resets' to a much smaller value at $|n|=n_s/2$. The Hall density beyond these points behaves as if the charge carriers contributing to transport are just those added beyond $|n|=n_s/2$ and roughly follow $n_H=n+n_s/2$ for $n<-n_s/2$ and $n_H=n-n_s/2$ for $n>n_s/2$. This behaviour is in agreement with the quantum oscillation frequency measurements shown in Fig. 5d.

This `resetting' effect is quickly suppressed by raising the temperature to about \SI{10}{\kelvin}. Beyond this temperature the Hall density shows a monotonic increase towards the band edge. We do note that at higher temperatures, the Hall density in the flat bands no longer follows $n_H=n$. This can possibly be explained by the thermal energy $kT$ being close to the bandwidth of the flat bands, and the Hall coefficient must take into consideration the contributions from thermally excited carriers into the higher energy highly dispersive bands, with opposite polarity. In contrast, the Hall density measured at very high densities when $|n|>n_s$ shows very linear behaviour according to $|n_H|=|n|-n_s$ regardless of the temperature up to \SI{30}{\kelvin}, which is consistent with the highly-dispersive, low-mass bands above and below the flat bands, as can be seen in Fig. 1c.

\end{document}